\documentclass[12pt,amstex,epsfig,fleqn]{article}
\usepackage{graphicx}
\usepackage{rotating}
\usepackage{subfigure}
\usepackage{hyperref}
\usepackage{amsmath}
\usepackage{enumerate}
\def\lesssim{\mathrel{\hbox{\rlap{\hbox{\lower5pt\hbox{$\sim$}}}\hbox{$<$}}}}
\def\gtrsim{\mathrel{\hbox{\rlap{\hbox{\lower5pt\hbox{$\sim$}}}\hbox{$>$}}}}

\def\mhbarsq{\bar{m}_{H}^2}
\def\rinv{R^{-1}}
\def\b1{B_1}
\def\h1pm{H_1^\pm}

\def\leps{\ell_1^S} 
 
\def\nutau1{\nu_{{tau}_1}}
\def\mb1{m_{_{\b1}}}
\def\mh1pm{m_{{H_1^\pm}}}

\def\mleps{m_{_{\ell_1^S}}} 
\def\mlepd{m_{_{\ell_1^D}}} 

\def\beq{\begin{equation}}   %
\def\eeq{\end{equation}}   %

\begin{document}

\begin{flushright}
   {\bf \small 
        HRI-P-09-09-002 \\
        RECAPP-HRI-2009-018 \\
        CU-PHYSICS/06-2009}
\end{flushright}

\vskip 30pt

\begin{center}
{\large \bf  Search for Higgs bosons of the Universal Extra Dimensions\\ 
at the Large Hadron Collider}

\vskip 20pt
Priyotosh Bandyopadhyay$^{a}$\footnote{priyotosh@hri.res.in},
Biplob Bhattacherjee$^{b}$\footnote{biplob@gmail.com},
AseshKrishna Datta$^a$\footnote{asesh@hri.res.in}
\vskip 20pt
{$^a$ \emph{Regional Centre for Accelerator-based Particle Physics 
(RECAPP) \\
Harish-Chandra Research Institute
\footnote{A constituent institute of Homi Bhabha National Institute (HBNI), 
Department of Atomic Energy, Government of India}  \\
Chhatnag Road, Jhunsi, Allahabad, India 211019} }\\
\vskip 5pt
{$^b$ \emph{Department of Physics \\
University of Calcutta, 92 A.P.C. Road, Kolkata, India 700009}}
\end{center}

\abstract{The Higgs sector of the Universal Extra Dimensions (UED) has a
rather involved setup. With one extra space dimension, the main ingredients 
to the construct are the higher Kaluza-Klein (KK) excitations
of the Standard Model Higgs boson and the fifth components of the
gauge fields which on compactification appear as scalar degrees of
freedom and can mix with the former thus leading to physical
KK-Higgs states of the scenario. In this work, we explore in 
detail the phenomenology of such a Higgs sector of the UED 
with the Large Hadron Collider (LHC) in focus. We work out 
relevant decay branching fractions involving the KK-Higgs 
excitations.  Possible production modes of the KK-Higgs bosons 
are then discussed with an emphasis on their associated production
with the third generation KK-quarks and that under the cascade 
decays of strongly interacting UED excitations which turn out 
to be the only phenomenologically significant modes. It is 
pointed out that the collider searches of such Higgs bosons face 
generic hardship due to soft end-products which result from 
severe degeneracies in the masses of the involved excitations in 
the minimal version of the UED (MUED). 
Generic implications of 
either observing some or all of the KK-Higgs bosons at the LHC are 
discussed.} 
%
\section{Introduction}
In recent times, the Universal Extra Dimensions (UED) proposed
by Applequist, Cheng and Dobrescu (the ACD framework)
\cite{Appelquist:2000nn} 
has emerged as a viable option for physics beyond the Standard 
Model (SM) (see also \cite{Antoniadis:1990ew} for earlier 
related ideas).
In the UED, 
the extra space
dimensions (the \emph{bulk}) are available 
\emph{universally} to
all the SM particles to propagate in, \emph{viz.}, the fermions, 
the gauge bosons and the only scalar (the Higgs boson) of the SM. 
The simplest incarnation of such a scenario has only one extra
space dimension (i.e., a 4+1 dimensional space-time) and is 
popularly known in the literature as the \emph{minimal} UED 
(MUED) framework. With the Large Hadron Collider about to begin
its operation,
the scenario attracted a lot of attention in terms of its 
discovery and distinction from an extremely popular alternative 
in the form of supersymmetry (SUSY) which could potentially 
fake the observations.

It is well-known that in a theory with odd number of space-time 
dimensions (like the case of MUED with 5 space-time dimensions), 
one cannot have chiral fermions, which is an observational fact. 
Compactification of the extra space dimension on an orbifold 
($S^1/Z_2$, in the present case) results in chiral fermions in 
the effective 4-dimensional theory \cite{Appelquist:2000nn}. 
Also, only four of the five components of the gauge bosons 
survive in the low energy theory. Thus, orbifold 
compactifications opens up the avenue of identifying the SM as 
the low-energy limit of a TeV-scale extra dimensional theory.

Compactification of the extra space dimension leads to infinite
towers of excitations (the Kaluza-Klein (KK) excitations) in the
effective 4-D theory for the fields which originally have access
to it (the bulk).  By construct, thus, all the SM particles
(quarks, leptons, gauge-bosons and Higgs) have their
KK-excitations in the effective 4-D theory. Also, in the process,
other towers of KK-excitations emerge which do not have their SM
counterparts. The presence of KK-towers for charged and neutral
CP-odd Higgs bosons provides a concrete example of this.
The phenomenology of KK-excitations has been studied in great
detail in recent times which covers its implications at
colliders
\cite{Rizzo:2001sd,Macesanu:2002db,Cheng:2002iz, Cheng:2002ab,
Bhattacharyya:2005vm, Battaglia:2005zf,Riemann:2005es,
Bhattacherjee:2005qe,Datta:2005zs,Macesanu:2005jx,Hooper:2007qk,
Bhattacherjee:2008ik,Matsumoto:2009tb,Bhattacharyya:2009br}, 
in electroweak/flavour sector via various low-energy observables
\cite{Agashe:2001xt,Agashe:2001ra,Appelquist:2001jz,
Chakraverty:2002qk,Buras:2002ej,Oliver:2002up,Buras:2003mk,
Khalil:2004qk,Bucci:2003fk,Bucci:2004xw,
Gogoladze:2006br,Mohanta:2006ae} 
and for dark matter/cosmology 
\cite{Cheng:2002iz,Hooper:2007qk,
Dienes:1998vg,Cheng:2002ej, Servant:2002aq, Servant:2002hb,
Majumdar:2002mw, Majumdar:2003dj, Kakizaki:2005en,Kakizaki:2005uy,
Burnell:2005hm,Kong:2005hn,Flacke:2005hb,Kakizaki:2006dz}

In MUED there are only two extra free parameters when compared to 
the SM. These are the radius of compactification $R$ 
(or, alternately 
$R^{-1}$, the scale of compactification) and the cutoff scale
$\Lambda$ of the theory. The MUED parameter space also includes
the SM Higgs mass $m_{_{H_{SM}}}$. It is thus noteworthy that MUED 
is much more economic than even the most constrained version of SUSY
scenarios, like the minimal supergravity (mSUGRA) framework.

It is to be noted that the phenomenology of MUED turns out to be 
remarkably similar and thus can masquerade as SUSY \cite{Cheng:2002ab}. 
This is because of two basic reasons. First, both frameworks 
predict new excitations bearing the same gauge and global 
quantum numbers corresponding to the SM particles. 
Second, in each of these scenarios there is a conserved discrete
symmetry: the so-called $R$-parity in the SUSY case while its
counterpart in the UED scenario is the KK-parity, $K_P = (-1)^n$,
where $n$ stands for the $n$-th level excitation. Conservation of
KK-parity is very central to the phenomenology of MUED  and is
similar in its ramifications to $R$-parity of SUSY scenarios. 
Vigorous efforts are on 
\cite{Battaglia:2005zf,Datta:2005zs,Bhattacherjee:2008ik,
Smillie:2005ar,Datta:2005vx,ArkaniHamed:2005px,Meade:2006dw,
Athanasiou:2006ef,Athanasiou:2006hv,
Wang:2006hk,Hooper:2006xe,Choi:2006mr,ArkaniHamed:2007fw,
Wang:2008sw,Hubisz:2008gg,Belyaev:2008pk,Burns:2008cp,
Belanger:2008gy,Burns:2009zi,Ehrenfeld:2009rt} 
in 
chalking out strategies and their respective efficiencies to 
distinguish between these two contending scenarios under 
different situations. These efforts broadly exploits two salient 
differences between the two scenarios. While in the case of MUED 
the KK-excitations all have the same spin as their SM 
counterparts, for SUSY they differ by half. Also, MUED spectrum 
is comprised of a tower of discrete KK-excitations of the SM 
particles unlike the SUSY spectrum.

As does $R$-parity for SUSY, KK-parity forbids tree-level 
contribution to weak-scale processes. In some earlier works it was
reported that the electroweak precision data
require a lower bound of $R^{-1} \gtrsim 250-300$ GeV 
\cite{Appelquist:2000nn,Appelquist:2002wb}. This finding was later 
shown to be consistent with observations in the $B$-physics sector 
\cite{Agashe:2001xt} and with the experimental results on anomalous 
magnetic moments \cite{Agashe:2001ra,Appelquist:2001jz}. However, in 
later years, with more data and newer analyses it has been observed 
that the lower bound on $R^{-1}$ can surpass the earlier value and
can be as high as 700 GeV \cite{Gogoladze:2006br,Flacke:2005hb,
Haisch:2007vb} at 99\% confidence level. 
Also, 
similar to $R$-parity in SUSY scenarios, KK-parity ensures that the 
lightest $n=1$ 
KK-paricle (the LKP, with odd KK-parity) is an absolutely stable 
(like the lightest SUSY particle, the LSP) weakly interacting 
massive particle (WIMP) and is a good dark matter 
candidate that could provide the right amount of cosmological relic-density 
\cite{Servant:2002aq}\footnote{Usually,
the level 1 KK-excitation of the $U(1)$ gauge field $B$ of the SM,
$B_1$, turns out to be the LKP in the MUED scenario. 
The Weinberg angle at higher KK-levels being very small 
\cite{Cheng:2002iz} $B_1$ is sometimes identified as
the lightest KK-photon in the literature. In Ref.\cite{Cembranos:2006gt}, 
however, possibilities of other level-1 excitations like the
charged Higgs boson (see section 2) and the excited tau lepton 
becoming the LKP were discussed 
in some detail.}. 
Given the allowed
mass-scale as low as few hundred GeV and with a viable dark matter 
candidate in its spectrum, the scenario thus offers 
exciting signatures at future generation colliders, particularly 
at the LHC (see references [3]--[15]).

Interestingly enough, in contrast to a huge amount of recent and on-going
activities  in so many different aspects of the MUED-phenomenology, 
the collider aspects of the Higgs sector of the MUED has not been 
discussed in any detail until very recently. 
The MUED Higgs sector was discussed in \cite{Burnell:2005hm,Kong:2005hn}
in the context of estimating the relic abundance of KK-dark matter. 
In ref. \cite{Cembranos:2006gt} the authors explore some characteristic 
regions of the MUED parameter space and their general bearings on the 
MUED Higgs phenomenology including those at the colliders (with rather
characteristic and novel signatures). 
To the best of our knowledge, dedicated and somewhat detailed collider 
studies of the MUED Higgs bosons were only taken up in refs. 
\cite{Bhattacherjee:2006jb,Bhattacherjee:2007wy}\footnote{
Another set of literature 
\cite{Petriello:2002uu,Datta:2005ew,Rai:2005vy,Hsieh:2008jg}
discuss the virtual 
effects of the MUED states on Higgs physics at future colliders 
including the LHC and the extent of their detectability. These studies
exploit Higgs boson production via gluon fusion at the LHC and its decay
in the two-photon final state. It was observed in ref.\cite{Rai:2005vy} that
the uncertainties involved, when combined, could potentially
dilute the deviations from the SM to an insignificant level unless
one has a better control on this uncertainties. 
In ref.\cite{Hsieh:2008jg} it
is observed that these virtual effects of the MUED may cease to be 
perceptible at the same time when the MUED excitations become heavy 
enough to evade direct discovery at the LHC.}. 

In the present work, we study the cascade decays of the level-1 MUED 
excitation as one of the major sources for the KK level-1 Higgs bosons 
of the MUED with $B_1$ as the LKP. We also look at the dominant decay 
modes of the KK Higgs excitations thus identifying the final states in 
which they 
can be searched for at the LHC. \emph{Also, for the first time ever, the full 
mixing in the third generation level-1 fermion sector
is taken into account and applied to collider studies.}
The mixings are phenomenologically less significant in KK level-1 tau lepton 
and bottom quark
sectors when compared to the corresponding top-quark sector.

The paper is organized as follows. In section 2 we briefly outline the
MUED framework \emph{a la} ACD \cite{Appelquist:2000nn} with particular 
emphasis on its Higgs sector \cite{Buras:2002ej}. 
The spectrum of Higgs bosons and its
nature and dependencies which are crucial for their studies at the LHC
are discussed. In this work, we restrict ourselves to the physics of 
the level-1 charged Higgs boson of the MUED. 
In section 3 we describe the possible decays of the KK-Higgs
bosons under different circumstances and their implications at the colliders. 
Section 4 deals with different production mechanisms for the KK-Higgs 
bosons and identify the significant ones at the LHC. We also present
the results of our  numerical analysis in terms of the signal-strength and tentative
reach in $\rinv$. Section 5 summarizes with an outlook that
touches upon issues that emerge in the process.

\section{The Higgs sector of the minimal UED}
The structure of the fermion and the gauge boson sectors of the MUED
have been discussed in the literature in rather details 
\cite{Appelquist:2000nn,ued-writeup}. The
corresponding mass-spectra have also been discussed extensively. It is
also established that radiative corrections \cite{Cheng:2002iz} to the 
masses of these
KK-excitations play crucial roles in the phenomenology of the MUED by
lifting the degeneracy in their masses which otherwise, at tree level, go
as $nR^{-1}$ where $n$ stands for the concerned level in the tower of
KK-excitations.

However, as pointed out in the Introduction, the structure and properties
of the Higgs sector of the MUED and the consequent phenomenological
implications it bears have not received enough attention (except in
the references mentioned earlier). So much so that none of the popular
packages had yet included the Higgs sector in their MUED implementations 
\cite{ued-writeup,Pukhov:2004ca,ElKacimi:2009zj}
In the following
subsections we outline the basic construct of the MUED Higgs sector which 
involves the gauge boson sector in a very characteristic way. Also, the 
Higgs-phenomenology in the MUED involves the KK-fermions and the 
gauge bosons in a crucial manner. We would highlight them in appropriate 
contexts. We would then point out some features and interplays that make 
the Higgs sector of the MUED rather special.
 
\subsection{The basic construct}
In the MUED with (4+1) dimensions we are confronted with two 
well-known but fundamental problems.  First, we cannot obtain chiral 
fermions which we find in nature. Second, the 5th components of the 
(4+1)-dimensional gauge fields behave as some weird CP-odd scalars  
in (3+1) dimensions which we do not see in nature. In the framework of
MUED, in order to accommodate the chiral fermions in the theory 
and to project out the unwanted scalar degrees of freedom, one seeks 
an orbifold compactification of the kind $S^1/Z_2$ for the extra space 
dimension (say, denoted by $y$), i.e., compactifying it on a circle 
$S^1$  (a sphere of genus 1) about an orbifold  identified with two  
diametrically opposite fixed points  $y=0,\pi R$  (a $Z_2$ symmetry 
dubbed the KK-parity) along the direction $y$ \cite{Appelquist:2000nn}, 
$R$ being the radius 
of compactification.  A framework like MUED with a small ($R^{-1} 
\sim {\cal O} \mathrm{(TeV)}$) but flat extra dimension thus would 
likely to have interesting implications for the LHC and their studies 
have attracted significant attention in recent times as indicated in
section 1.

The boundary conditions imposed at the two orbifold fixed points 
determine the KK expansion for different fields. A scalar field has to
be either odd or even under the transformation ${\cal P}: y \to -y$
at $y=0, \pi R$. Thus, $\partial_5 \phi^+=0$ for the even fields
(the Neumann boundary condition) and $\phi^-=0$ for the odd fields
(the Dirichlet boundary condition) at the fixed points. The associated
Fourier expansions of (4+1)-dimensional scalar fields in terms of the 
(3+1)-dimensional KK-modes look like 
\begin{eqnarray}
\phi^+(x^\mu, y) &=& \frac{1}{\sqrt{2 \pi R}} \phi^+_{(0)}(x^\mu)
                            + \frac{1}{\sqrt{\pi R}} \sum_{n=1}^\infty
                               \phi^+_{(n)}(x^\mu) \cos \frac{ny}{R}  \nonumber \\
\phi^-(x^\mu, y) &=& \hskip 90pt \frac{1}{\sqrt{\pi R}} 
\sum_{n=1}^\infty  \phi^+_{(n)}(x^\mu) \sin \frac{ny}{R}
\label{scalarkkmodes}
\end{eqnarray}
where $\mu=0,1,2,3$ stands for the four ordinary non-compact space-time 
coordinates and $\phi^\pm_{(n)}(x^\mu)$ are the (3+1) dimensional
KK-modes with $n=0$ (level $0$) standing for the SM excitation. Clearly,
by construct and as required, a $Z_2$-odd state at level `0' is projected out
by the choice of boundary conditions. The odd degrees of freedom start
appearing only from level `1' and hence, they do not have any corresponding
SM field.

On the other hand, a vector field $A^M$ possesses 5 components in
(4+1) dimensions, i.e., M=0, 1, 2, 3 and 5, conventionally. The orbifold
compactification of the form  $\partial_5 A^\mu = 0$ along with
$A^5 = 0$ ensures that the first 4 components are even under 
${\cal P}_5$ while the 5-th component transforms as an odd degree of 
freedom. The Fourier expansions of the (4+1) dimensional vector field in
terms of corresponding (3+1) dimensional KK-modes are
\begin{eqnarray}
A^\mu(x^\mu, y) &=& \frac{1}{\sqrt{2 \pi R}} A^\mu_{(0)}(x^\mu)
                                     + \frac{1}{\sqrt{\pi R}} \sum_{n=1}^\infty
                           A^\mu_{(n)}(x^\mu) \cos \frac{ny}{R}  \nonumber \\
A_5(x^\mu, y) &=&  \hskip 92pt  \frac{1}{\sqrt{\pi R}} 
        \sum_{n=1}^\infty
                                         A^5_{(n)}(x^\mu) \sin \frac{ny}{R}
\label{gaugekkmodes}
\end{eqnarray}
As in the case for the scalars above, here also,  the $Z_2$-odd gauge degrees 
of freedom which behave as scalars in (3+1) dimensional description appear 
only from the first KK level and thus have no corresponding SM excitation.
These extra $Z_2$-odd scalar degrees of freedom in equations 
\ref{scalarkkmodes} and \ref{gaugekkmodes} may combine to give
new scalar degrees of freedom at level 1 onwards. 

Similarly, with appropriate boundary conditions imposed at the 
orbifold fixed-points the expansion of the the (4+1) dimensional 
4-component Dirac fermions  are
\begin{eqnarray}
\psi^+(x^\mu,y) &=& \frac{1}{\sqrt{2 \pi R}} \psi_{R(0)}(x^\mu)
          + \frac{1}{\pi R} \sum_{n=1}^\infty 
\left (  \psi_{R(n)}(x^\mu) \cos \frac{ny}{R}
      +  \psi_{L(n)}(x^\mu) \sin \frac{ny}{R}
      \right ) \nonumber \\
\psi^-(x^\mu,y) &=& \frac{1}{\sqrt{2 \pi R}} \psi_{L(0)}(x^\mu)
          + \frac{1}{\pi R} \sum_{n=1}^\infty 
\left (  \psi_{L(n)}(x^\mu) \cos \frac{ny}{R}
      +  \psi_{R(n)}(x^\mu) \sin \frac{ny}{R}
      \right )
\label{fermionkkmodes}
\end{eqnarray}
As should be the case, the zero-modes, i.e., the SM fermions are either left or
right handed. Thus, the chiral fermions of the SM are projected in by the
orbifolding while all the excited KK fermions are made up of chiral pairs, i.e.,
these are vector-like fermions. In other words, while the SM fermions are
constructed out of chiral components which have different $SU(2)_L \times
U(1)_Y$ quantum numbers, in the MUED the components have the 
corresponding quantum numbers the same. This difference is 
phenomenologically crucial and shows up in the couplings of these vector
like states with the gauge bosons and the Higgs bosons of the MUED.

We now collect the components that build up the MUED Higgs sector.
Since in UED all SM excitations including the Higgs degrees of freedom 
can propagate in the 5-dimensional bulk, on compactification, these also 
get a tower of scalars. The KK excitations of the components of the SM Higgs
doublet can be expressed using the usual compact notation:
\begin{equation}
H_{(n)} = \left ( 
\begin{array}{cc}
i \chi^+_{(n)} \\
\frac{1}{\sqrt{2}} (h_{(n)} + i \chi_{(n)})
\end{array}
      \right )
\end{equation}
where $\chi^+_{(n)}$ are complex positively charged fields and hence
CP-odd in nature \underline{($Z_2$-odd as well)}.  On the other hand, 
starting KK level 1, we already have the 5th components of the electroweak 
gauge bosons which behave as CP($Z_2$)-odd scalars on compactification  
to (3+1) dimensions. These excitations are $B_{(n)}^5, W_{3(n)}^5$
and $W^{1,2(5)}_{(n)}$. Note that these scalar degrees of freedom are
not exact extra-dimensional analogues of the SM Nambu-Goldstone bosons
since they all receive the KK-mass of $n\over R$ because of the derivative
operator $\partial_5$ operating on them.  
When the electroweak 
symmetry is broken $B_{(n)}^5$ and $W_{3(n)}^5$ mix to give 
$\gamma^5_{(n)}$ and  $Z^5_{(n)}$ while $W^{1(5)}_{(n)}$ and 
$W^{2(5)}_{(n)}$ combine to result in  $W^{\pm (5)}_{(n)}$
analogous to what happens in the SM.

Now,  $Z^5_{(n)}$ mixes with  $\chi_{(n)}$ while $W^{\pm 5}_{(n)}$
mixes with $\chi^\pm_{(n)}$ to form not only the effective 
Nambu-Goldstone bosons $G^0_{(n)}$ and $G^\pm_{(n)}$ but also the
three physical Higgs bosons, the CP-odd neutral $A^0_{(n)}$ and the charged
Higgs bosons  $H^\pm_{(n)}$. The Goldstone states
$G^0_{(n)}$ and $G^\pm_{(n)}$ are absorbed by the SM gauge bosons
to become massive keeping the gauge invariance intact. The mutually 
orthogonal Goldstone and physical Higgs boson states are given by
\begin{eqnarray}
G^0_{(n)} = \frac{1}{M_{Z_{(n)}}} 
\left [M_Z \chi_{(n)} - \frac{n}{R} Z^5_{(n)} \right ]
\qquad \qquad A^0_{(n)} = \frac{1}{M_{Z_{(n)}}}
\left [\frac{n}{R} \chi_{(n)} + M_Z Z^5_{(n)} \right] \\
G^\pm_{(n)} = \frac{1}{M_{W_{(n)}}} 
\left [M_W \chi^\pm_{(n)} - \frac{n}{R} W^{\pm 5}_{(n)} \right ]
\qquad \qquad A^\pm_{(n)} = \frac{1}{M_{W_{(n)}}}
\left [\frac{n}{R} \chi^\pm _{(n)} + M_W W^{\pm 5}_{(n)} \right]
\label{higgs-goldstones}
\end{eqnarray}
where $M_{Z_{(n)}}$ and $M_{W_{(n)}}$ are given by 
\[ M_{Z_{(n)}}^2 = \frac{n^2}{R^2} + M_Z^2 
\qquad \quad 
 M_{W_{(n)}}^2 = \frac{n^2}{R^2} + M_W^2
\]
Also, note that the KK excitations of the SM Higgs boson is physical CP-even
states which do not mix with the CP-odd {$\gamma^5_{(n)}$ states.
The latter remain to be additional unphysical scalar modes which act as
Goldstone bosons absorbing which $\gamma^\mu_{(n)}$ become massive
for $n \geq 1$. As implied  by equation \ref{gaugekkmodes}, $\gamma^5$
would not have `0' modes which thus keeps the usual photon (level `0')
massless.

It is evident from equation \ref{higgs-goldstones} that with increasing `$n$',
i.e., at higher KK-levels, the Goldstone modes are dominated by 
$Z^5_{(n)}$ and $W^{\pm 5}_{(n)}$  while the physical Higgs modes
are mostly $\chi_{(n)}$ and $\chi^\pm_{(n)}$. Taking a cue from
equation \ref{scalarkkmodes} it is extremely crucial to note that the Higgs
states $A_{(n)}$ and $H^\pm_{(n)}$ do not have a `0' mode. They first
appear at the KK level 1. To summarise, there are 4 new Higgs states at KK 
level 1 which are $A_{(n)}$, $H^\pm_{(n)}$ and $H_{(n)}$ of which only
the last one has a level `0' mate which is the usual SM Higgs boson.
Thus, if we restrict ourselves to KK level-1 for studying the Higgs bosons, 
we end up with 5 Higgs bosons in total of which only one (the level `0' (SM)
Higgs boson) has even KK-parity while the other four from level 1 are of odd 
KK-parity. 

The interesting thing to note here is that the number of Higgs bosons (up to 
KK level 1) in the MUED framework is exactly the same as that in the Minimal
Supersymmetric Standard Model (MSSM). Thus, much talked about faking
of the MSSM by a MUED like scenario encompassing the fermion and the 
gauge  boson sectors, comes to a full circle with the similarity in the scalar 
sectors that unfolds. The origins of the scalar states in the two scenarios, 
however, are in clear contrast. The five Higgs bosons of the MSSM are the 
well-known outcome of a generic scenario with two Higgs doublets. They 
all are $R$-parity even. In contrast, in the MUED, the only (SM) Higgs doublet 
suffices. There, the multiplicity of the Higgs states is traced back to the 
subtle presence of KK towers for the additional (scalar) degrees of freedom 
pertaining to the gauge sector that result from orbifolding. Also, only one 
of these Higgs bosons (the level-`0' or the SM one) carries even KK-parity 
while the other four are  odd under the same. This feature is particularly
crucial for the phenomenology of the Higgs bosons in the MUED where
KK-parity is conserved. It is not at all unexpected if we remember what role 
$R$-parity conservation plays in shaping the  SUSY phenomenology.

\subsection{The mass spectrum and the mass-splittings}
In this subsection we discuss the Higgs mass spectrum of the MUED  vis-a-vis
the spectrum of other MUED excitations and reflect on their phenomenological 
implications.

The Higgs spectrum of the MUED at KK level $n$ is determined by the 
following relations:
\begin{eqnarray}
m_{H_n^0} &=& m_n^2 + m_H^2 +\hat{\delta}m_{H_n}^2   
\nonumber \\
m_{A_n^0} &=& m_n^2 + m_Z^2 +\hat{\delta}m_{H_n}^2
\nonumber \\
m_{H^\pm_n} &=& m_n^2 + m_W^2 +\hat{\delta}m_{H_n}^2
\label{mass-kkhiggs}
\end{eqnarray}
where, $m_H$ is the mass of the SM Higgs boson and 
$\hat{\delta}m_{H_n}^2$ represents the total one-loop correction
(including both bulk and boundary contributions, the former being identically 
equal to zero in the case of Higgs bosons) which is universal to all the Higgs
states at level 1 and is given by \cite{Cheng:2002iz}
\begin{equation}
\hat{\delta}m_{H_n}^2 = m_n^2 
\left ( 
\frac{3}{2} g_2^2 + \frac{3}{4} g'^2 - \lambda_H
\right)
\frac{1}{16 \pi^2} \ln \frac{\Lambda^2}{\mu^2} + \bar{m}_{H}^2
\label{masscorrection-kkhiggs}
\end{equation}
where $m_n$ refers to the universal KK mass arising out of compactification
at level $n$ and given by $\frac{n}{R}$, $g_2 $ and $g'$ stand for the 
$SU(2)$ and the $U(1)$ gauge couplings respectively, $\lambda_H$
represents the standard Higgs quartic coupling and is given by
$\lambda_H=m_H^2/v^2$, $\Lambda$ gives the cut-off
scale (${\cal{O}}$(TeV)) up to which the UED description is valid, $\mu$
stands for the renormalization scale and $\bar{m}_{H}^2$ is the universal
boundary mass term for the Higgs mode which is a free parameter set to zero 
in the MUED but we shall also look into the implications of non-zero values
for the same. Thus, the Higgs sector of the MUED is completely specified by 
the following free parameters: $\Lambda$, $R^{-1}$, $m_H$, $\mu$ and
$\bar{m}_{H}^2$.

From equations \ref{mass-kkhiggs} and \ref{masscorrection-kkhiggs} it is 
apparent that the hierarchy of the 4 Higgs boson masses at the first  KK level 
is as follows:
\begin{equation}
m_{H_1^\pm} < m_{A^0_1} < m_{H^0_1}
\end{equation}
It is also rather clear that the terms responsible for the mass-splittings among
the KK-Higgs masses are those involving the masses of the SM gauge bosons 
and the SM Higgs boson. $M_W$ and $M_Z$ being the experimentally
measured quantities, the individual masses of the level-1 charged and the 
CP-odd neutral Higgs bosons (as well as their mutual splittings) are pretty much 
determined. $m_{H^0_1}$, however, remains a function of  the unknown
SM Higgs boson mass but guaranteed to be the heaviest of them all with  $m_H$
(i.e., the KK level-0 Higgs boson which is the SM Higgs boson) satisfying the 
LEP bound (i.e., $m_H \gtrsim 114.4$ GeV) \cite{Barate:2003sz}. 
It is interesting to note that with increasing $m_H$, $\lambda_H$ increases
(in equation \ref{masscorrection-kkhiggs})
thus decreasing $\hat{\delta}m_{H_n^2}$ which results in decreasing
$m_{H_1^0}$. Also, with 
increasing $R^{-1}$, the masses of all the  KK Higgs bosons get dominated 
by the KK-mass term ($n/R$) and the mass-splittings among the excited 
Higgs bosons become increasingly small. However, phenomenologically,
the more relevant parameters are the splittings among the KK Higgs bosons and
the low-lying KK excitations like the singlet KK leptons and the lightest KK 
particle (LKP) in which the former could decay. In section 3, we illustrate
a representative situation and its phenomenological implications.

\section{Decay of the level-1 charged Higgs boson}
%
In this section, we discuss the possible decay modes of the level-1
charged Higgs boson
and the expected features in the kinematics of its decay products. It is clear
from the last section  that the KK Higgs bosons are some of the lowest
lying states in the MUED spectrum. With $\mhbarsq$ set to zero, only $B_1$
(the LKP), the KK leptons and the KK neutrinos can become lighter than the 
level-1 charged KK Higgs boson ($\h1pm$) depending upon the value of $\rinv$. 
However, for larger $\rinv$ (as would be demonstrated later in this section)
only singlet type KK-leptons could become lighter
than $\h1pm$ and thus, are the states to which the latter can 
favourably decay into.

There are thus a few characteristic modes in which $\h1pm$ can decay
depending upon the spectrum of the above mentioned excitations.  In the MUED, 
with varying input parameters (in particular, $\rinv$), there can be the 
following two phenomenologically relevant mass-orderings involving the 
above-mentioned low-lying MUED excitations which are somewhat degenerate
in nature:
\begin{enumerate}[(i)]
\item \underline{$\mb1 < \mleps  < \mh1pm < \mlepd$}: In this case, 
$\h1pm$ would undergo the two-body decay $\h1pm \to \leps \, \nu_\ell$ 
followed by $\leps \to \ell \b1$. The hardness of the $\ell$ is limited
by and increases with the mass-splitting ($\Delta m_{_{(\leps \b1)}}$) 
between 
$\leps$ and $\b1$. From Fig. \ref{fig:mass-split} it is clear that the 
mass-split in reference and hence the hardness of the lepton 
would increase 
with increasing $\rinv$. Thus, for $500 \, \mathrm{GeV} < \rinv < 2000$ GeV, 
one can expect
$p_T^\ell$ to range between \underline{$15 \, \mathrm{GeV} < p_T^\ell < 25$ GeV}.
Detection of such a soft lepton would be a challenge at the LHC.
Nevertheless, uncovering scenarios with such generic degeneracies in the 
spectrum would necessarily require such a resolution\footnote{A similar
situation occurs in a supersymmetric scenario in the so-called stau
coannihilation region where the stau and the LSP (the lightest neutralino)
are rather degenerate \cite{Guchait:2008ar,Ball:2007zza}.}.  
\item \underline{$\mb1  < \mh1pm < \mleps < \mlepd$}:
Here, $\h1pm$ is the next-to-lightest KK particle (NLKP).\footnote{Phase 2 of
the phase diagram in ref.\cite{Cembranos:2006gt}.}
$\h1pm$ could only undergo  3-body decays of the form 
$\h1pm \stackrel{q_1^*,\ell_1^*}{\longrightarrow} f \bar{f}^\prime \b1$
via off-shell quarks and leptons. 
It has earlier been observed in ref.\cite{Cembranos:2006gt} that the decay
lengths can be as big as 20 cm for $\Delta m = 1$ GeV and for 
$\Delta m \lesssim 0.4$ GeV $\h1pm$ becomes essentially stable at colliders.
These situations could result in displaced vertices and non-zero impact parameters, 
signals of slow and metastable charged with highly ionizing tracks etc.
However, not the
individual but the combined hardness of the SM fermions (quarks or lepton and 
neutrino) in the
final state is governed by the mass-splitting ($\Delta m_{_{(\h1pm \b1)}}$) 
between $\mh1pm$ and $\b1$. This mutual sharing would render the final-state
fermions even softer, especially when the mass splitting is already miniscule, 
as is the case here. It will be a challenge at the LHC (or, for that matter, at any
future collider) to detect such soft fermions and hence the signatures that depend
critically on this.   
\end{enumerate}
\begin{figure}[htbp]
\centering
\includegraphics[height=8.5cm, width= 6.0cm, angle =270]{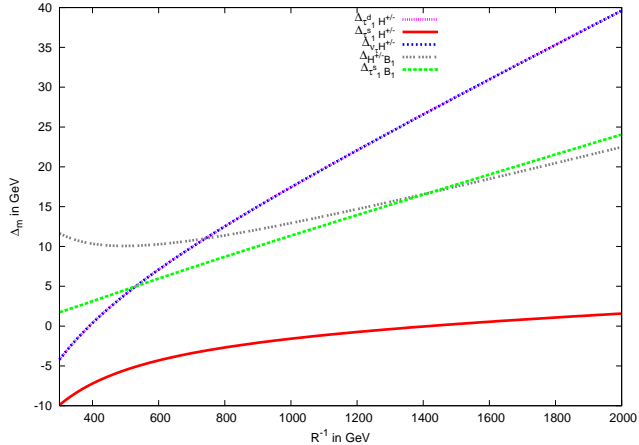}
\caption{\small Variation of the mass-splittings between different low-lying 
level-1 KK-excitations of the MUED as functions of $R^{-1}$ and for
$\Lambda R=20$ and $m_{H_{_{SM}}}=120$ GeV. Convention followed for 
the legend is that the mass of the second particle is subtracted from that
of the first.}
\label{fig:mass-split}
\end{figure}
%
Some crucial observations worth special mentions:
\begin{itemize}
\item First, although the 
fermion detection efficiencies would increase with increasing $\rinv$ 
(at least for extreme to moderately soft fermions),
the optimal reach in the latter via  search for $\h1pm$ would be 
determined by how sharply the production cross section of $\h1pm$ drops with
$\rinv$. 
\item Second, in the two cases discussed above, the degeneracy of the
mother and the daughter (KK) excitations are such that in the first case
the MUED excitations would decay promptly while the latter one is likely
to have exotic signals from long-lived, metastable excitations. The region
of parameter space where such phenomenon may occur depends crucially on
the mass of the SM Higgs boson \cite{Cembranos:2006gt}. In our analysis,
however, we have kept $m_{H_{_{SM}}}$ fixed at 120 GeV for which the latter phase
takes over at around $\rinv=1400$ GeV (see the bottom-most curve of 
Fig. \ref{fig:mass-split}. 
\item Lastly, the LKP, though by virtue of being rather heavy ($\mb1 \sim \rinv$) 
could  carry a lot of  $p_T$,
the vector-sum of the same for two of them\footnote{Note that, similar to SUSY
scenarios with conserved $R$-parity, KK-parity in MUED ensures that level-1
MUED states would be pair-produced at the LHC and they would ultimately cascade
to a pair of LKPs ($\b1$, in our case).} in the final state may suffer
significant cancellation because of the afore-mentioned mass-degeneracy 
between the decaying particle and the the LKP\footnote{Similar observations 
were made in Refs. \cite{Kawagoe:2006sm,Strassler:2008jq}.}. Thus, the 
missing $p_T$ in any event turns out to be characteristically soft in 
contrary to the generic expectation. These are the kinematic regimes where 
the backgrounds are generally very high thus making the hunt for exotics 
like $\h1pm$ further challenging.  A detailed simulation with major 
detector effects incorporated would be required for a reasonable 
understanding of the phenomenon.
\end{itemize}

\begin{figure}[htbp]
\centering
\includegraphics[height=8.5cm, width= 6.0cm, angle =270]{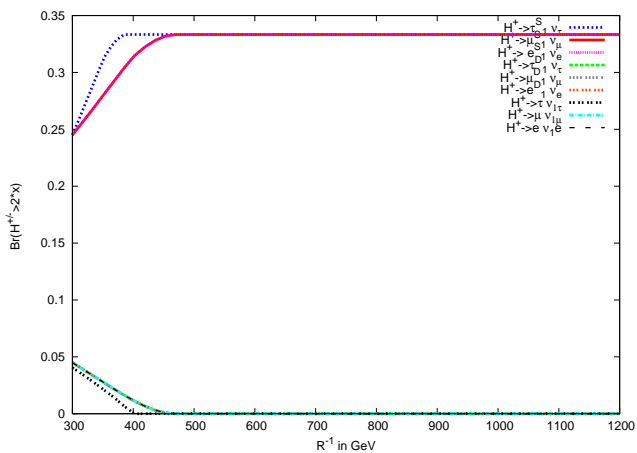}
\caption{\small Variation of the decay branching fractions 
of the level-1
charged Higgs boson ($\h1pm$) to leptons and neutrinos
as a function of $\rinv$ (in GeV). 
The blue 
(top) curve stands for the combined branching fraction of $\h1pm$ to
singlet level-1 KK leptons
while the red and black (bottom) correspond to the combined fractions
to doublet level-1 KK leptons and level-1 KK-neutrinos respectively.}
\label{fig:kkhiggsdk}
\end{figure}
%
Consequently, searches for the KK Higgs bosons are supposed to be heavily 
plagued by the background thus making their observation extremely difficult
at any future collider including the LHC. In Fig. \ref{fig:kkhiggsdk} 
we present the branching fractions of the charged level-1 KK-Higgs boson 
to level-1 KK leptons and KK neutrinos.
The behaviour of the curves can be explained with the help of the expressions
for the respective couplings as given in ref.\cite{Buras:2002ej}. 
However, the interplay of
different parameters that are instrumental is quite complex. One can, however,
try to follow the proceedings in a somewhat intuitive way as follows.

Note that the couplings of a pure KK Higgs boson to the KK leptons and/or KK 
neutrinos necessarily involve a chirality-flip. Since, a neutrino (or its KK
excitation) is a helicity eigenstate (left-chiral), its interaction vertex 
with a `pure' charged Higgs boson could only involve a right chiral charged 
KK lepton (or its SM counterpart) whose strength is proportional to the lepton 
Yukawa coupling and is, in turn, proportional to its mass. This issue is clear 
if we note that the decay branching fraction to $\tau_1^S$ is a little more 
enhanced that that for the other two lepton flavours for low $\rinv$.   

The overwhelmingly large branching fraction to singlet
KK-leptons is due to the availability of phase space 
while the same gets closed
for $\h1pm \to \ell^D_1 \nu$ for $\rinv \gtrsim 400$ GeV since
$m_{\ell_1^D} > m_{\h1pm}$ for such $\rinv$ (see Fig. \ref{fig:mass-split}).
Once they lead, the individual branching fractions to $e_1^S$, $\mu_1^S$
and $\tau_1^S$ are mainly governed by the term proportional to the respective 
fermion masses for low $\rinv$.
It should be remembered that the absolute width for decays of $\h1pm$ in
these modes are still small ($\lesssim {\cal O}(10^{-5}$ GeV)) though ensures promptness
of its decay. With growing $\rinv$ the terms in the coupling conspire to
ensure a universality among the three lepton flavours in the decay of $\h1pm$.
This feature goes a long way in shaping the phenomenology at the LHC. In regions
of MUED parameter space where $\ell_1^S$-s are the NLKP $\h1pm$ decays almost
indiscriminately to all three flavours. With $p_T^\ell$ being governed by
the mass-split $\Delta m_{\ell_1^S \b1}$  (from the subsequent decays  of
$\ell_1^S \to \ell \b1$) it is expected (from Fig. \ref{fig:mass-split}) 
that the former can attain a value of around 15 GeV before a new phase emerges
at around 1.4 TeV. The important point to note here is that one may try to
tag the soft electrons and muons which is comparatively easier instead of
looking for such a soft tau.

~However, as the charged Higgs boson 
(as well as all the excited Higgs bosons) has an admixture of 5th (scalar) 
component of the 5-dimensional gauge bosons, this may lead to some  
chirality-conserving (gauge) couplings albeit suppressed by the mass of 
the appropriate KK gauge boson. This feature is clear from the fact that
there are some small but non-zero branching fractions to doublet KK-leptons
and to KK-neutrinos at relatively small $\rinv$ which quickly 
(at around $\rinv=400$ GeV) drops to insignificant levels as $\rinv$ grows. 

In this context, it should be mentioned that with increasing mass of the
SM Higgs boson, $\h1pm$ may become the LKP for larger values of $\rinv$.
Such a phase raises the standard debates pertaining 
to the implications of a stable charged particle as the dark matter candidate 
and a clarification to this is given in \cite{Cembranos:2006gt}.
We, however, do not get into the analysis of this 
phase at this stage.
%
\section{Production of level-1 KK Higgs bosons}

It is rather obvious from the previous section that finding the KK 
Higgs bosons at colliders would be a difficult task due to the kinematic 
features (softness) their decay products bear. Thus, any serious attempt 
to resurrect the possibility of finding the KK Higgs boson should, in the 
first place, ensure that the production cross section of them are 
reasonably high so as to make up for a low overall efficiency that would
be plaguing 
their detection. Hence, in this section, we take up a brief study on the 
production rates of the KK Higgs bosons (particularly, for the charged 
KK Higgs boson).

Level-1 KK Higgs bosons have odd KK-parity and hence cannot be produced 
as a single resonance unlike the Higgs bosons of the SM or those of the 
SUSY scenarios which have even $R$-parity. KK Higgs bosons can only be 
produced in association with another level-1 KK excitation which has a 
similar mass as the KK Higgs boson ($\sim \rinv$. This brings in a 
significant demand 
on the available phase space when compared to the cases of Higgs boson
production in SM and the SUSY scenarios where not only their resonant 
productions but also their associated productions with lighter particles 
like the SM gauge bosons and fermions are promising. 

The production processes of the level-1 KK Higgs bosons  can be broadly
divided into two categories: (i) direct production processes and 
(ii) production under UED cascade decays. The direct production 
processes include pair production of the level-1 Higgs bosons and their 
production in association with level-1 gauge bosons like $Z_1$ and 
$W^\pm_1$. It is already well-known 
that the rates for these processes are not at all significant at the LHC. 
This is not unexpected since these involve production of two massive 
$\sim \rinv$ particles mainly driven by weak interaction. We leave 
the detail 
exploration of such processes for a future work. 
Pair production of $t_1\bar{t}_1$ followed by their decays to charged 
level-1 KK-Higgs bosons 
was discussed in Ref. \cite{Bhattacherjee:2007wy}.
In the present work, we take a special note of the couplings involved 
in the latter processes and concentrate on the production of the 
level-1 charged KK-Higgs bosons under cascades (involving UED excitations) 
at the LHC. This is in line with and motivated by similar studies 
undertaken for the MSSM Higgs bosons in recent times and which turned 
out to be very promising\footnote{This is a 
possibility as it is well-known that corresponding to each cascade 
diagram involving SUSY particles in the MSSM there is a corresponding 
``twin diagram'' \cite{Datta:2005zs} involving their KK-counterparts in MUED.
While for the MSSM Higgs bosons these 
may not serve as the discovery channels but would play complementary 
and/or supplementary roles in deciphering the MSSM Higgs system, in case 
of  the level-1 KK Higgs bosons these turn out to be the only modes that 
are phenomenologically interesting.} \cite{Datta:2001qs,Datta:2003iz}. 

\subsection{Production of $H^\pm_1$ under UED cascades}
The production of $H^\pm_1$ under UED cascades takes advantage of the
large production cross section of the strongly interacting UED excitations 
(level-1 KK gluon and  quarks) at the LHC. The yield of $H^\pm_1$ under 
cascade decays of these excitations depends crucially on how favourably they
decay to $H^\pm_1$.
To this effect, pair-production of level-1 KK gluon ($pp \to g_1 g_1$) and the
associated productions of level-1 gluon and level-1 quarks ($pp \to g_1 q_1$)
play the dominant roles. For these, it is the effective decay branching 
fraction of $g_1$ to $H^\pm_1$ that holds the key. Decays of $q_1$ to $H^\pm_1$
are suppressed for the first two generations of quarks and 
are only significant for the ones from the third generation, i.e., for $b_1$ 
and $t_1$.\footnote{See, for example, Appendix A, equations A.3, A.4 and A.33 
to A.36. of ref.\cite{Buras:2002ej}.} Thus, pair productions of the third 
generation level-1 KK quarks 
have somewhat significant contributions to $H^\pm_1$ through their decays.
In fact, to obtain $H^\pm_1$ from the decay of $g_1$, the cascades almost 
necessarily have to go via an intermediate state involving a third generation 
level-1 quark, i.e., $g_1 \to b(t) \, \bar{b}_1(\bar{t}_1) \to H^\pm_1 + X$.
Hence, an important part of the present work is the study of the decays of
level-1 gluon to $b_1$ and $t_1$ and their subsequent decays to $H^\pm_1$.
We reiterate that ($SU(2)$ and $U(1)$) mixing in the third generation quark 
and lepton sector has been fully incorporated in our present
analysis. It is observed that the lighter mass eigenstates
$t_1$ and $b_1$ have predominantly the $SU(2)$ component while 
the heavier $t_2$ and $b_2$ states are mostly of $U(1)$ type. 
These relative strengths of different couplings play crucial 
roles in subsequent analysis.

The couplings that are instrumental in determining the
strengths of different branching modes of $t^{(1,2)}_1$ and
$b^{(1,2)}_1$ to $H^\pm_1$ are $ H_1^+ ~ \bar{t}_1 \, b$,
$H_1^+~\overline t_2 \, b$ , $H_1^-~ \overline b_1 \,t$,
$H_1^-~ \overline b_2 \, t$ (and their hermitian conjugates).
We checked and used these couplings as given in ref. 
\cite{Buras:2002ej}.
A minute inspection of these couplings reveals that the $U(1)$
type (heavier) level-1 top quark has a coupling with $H_1^\pm$ 
which is similar to the coupling involving the $SU(2)$ type 
(lighter) level-1 KK bottom quark and $H_1^\pm$. These couplings
are enhanced compared to the other set of similar couplings
relevant in the present analysis, i.e., the couplings of 
$H_1^\pm$ with lighter level-1 KK top quark and that with the 
heavier level-1 KK bottom quark. As we would see, these 
relative strengths of different couplings play crucial roles 
in subsequent analysis.

For numerical purposes we use CalcHEP \cite{Pukhov:2004ca} and 
modified the available model files \cite{ued-writeup}
for MUED to incorporate the Higgs sector up to KK level-1 by including the
couplings of level-1 Higgs bosons with different SM and level-1 UED 
excitations \cite{Prv-Comm-Kribs-Kong}. To calculate the cross sections
for different processes we make use of the CTEQ6L \cite{Pumplin:2002vw} parton
distribution functions with the renormalization/factorization scale set at
$\sqrt{\hat{s}}$. The present study is carried out at the parton-level
to establish that cascade decays of level-1 KK particles of the MUED
could turn out to be healthy sources for the (level-1) charged Higgs boson
at the LHC that may aid their search.

\subsection{Productions and decays: modes and rates}
Guided by the discussion above,
in this subsection we analyze the production and decay schemes of different level-1
KK excitations that would have crucial bearing on the yield of $\h1pm$ at the
LHC under UED cascades.   
\begin{figure}[htbp]
\centering
\subfigure[]
{\label{fig:strong-prod}
\includegraphics[height=7.5cm, width= 5.5cm, angle =270]{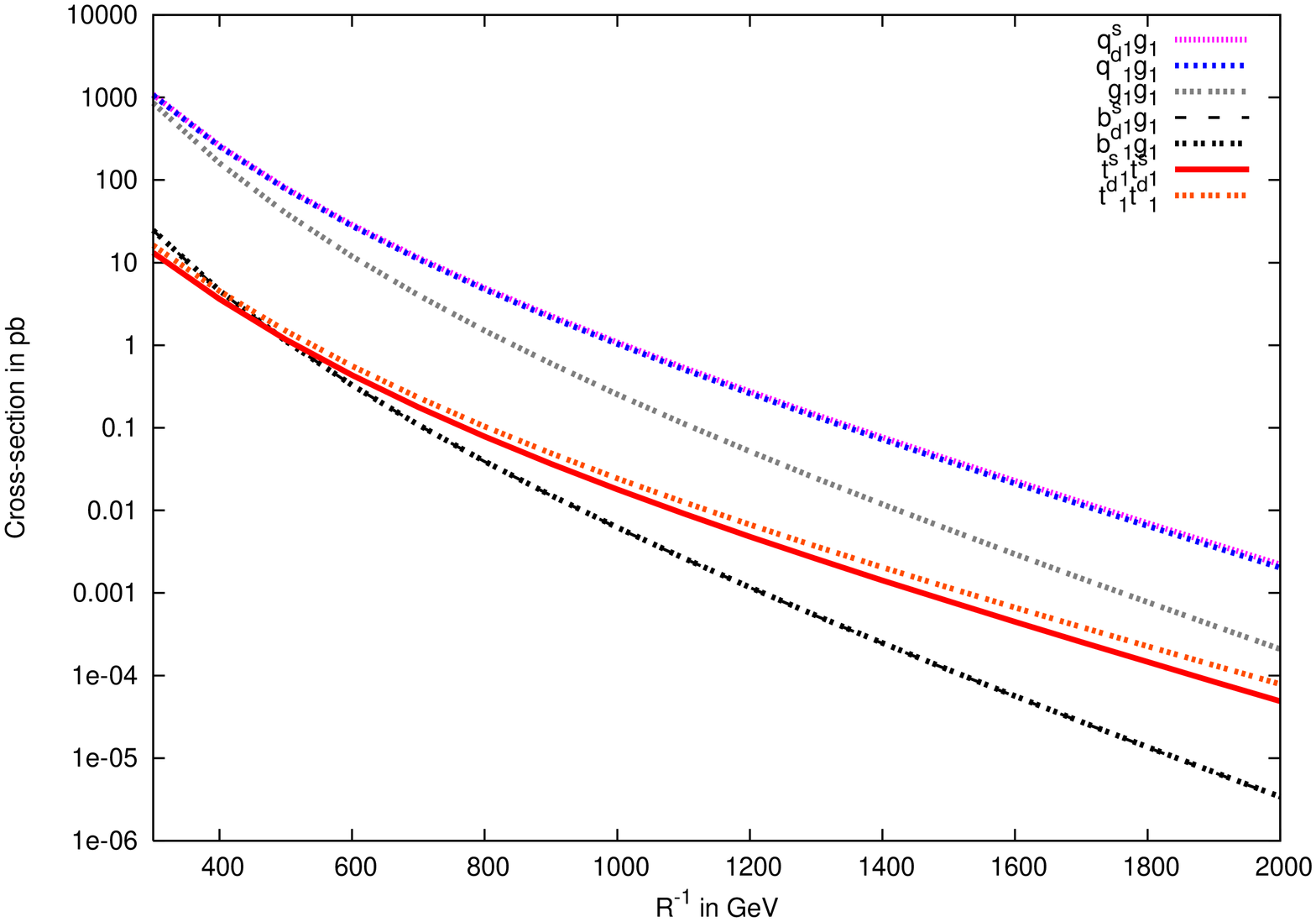}}
\hskip 15pt
\subfigure[]
{\label{fig:gluon1-decay}
\includegraphics[height=7.5cm, width= 5.5cm, angle =270]{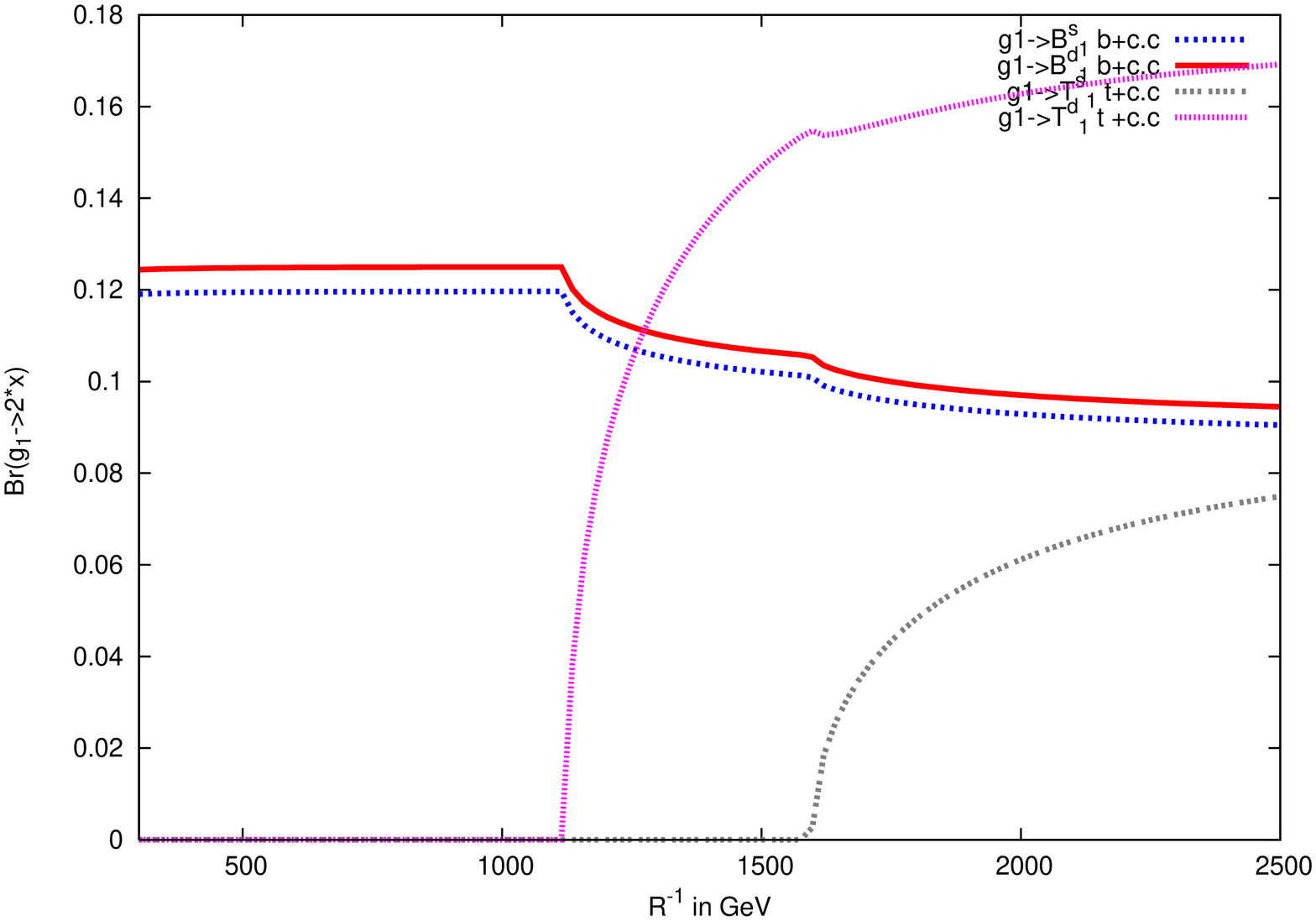}}
\caption{\small
(a) Variations of different strong production cross sections at an LHC
energy of 14 TeV. $q_1$ in the legend stands for the level-1 KK quarks
from the first two generations. Contributions from the charge-conjugated 
final states are taken into account wherever relevant.
(b) Variation of the decay branching fractions 
of level-1 KK-gluon
to level-1 third generation KK-quarks as functions of $R^{-1}$.
}
\end{figure}

In Fig. \ref{fig:strong-prod} we illustrate the basic strong production cross 
sections for $g_1$-pair, $g_1 q_1$ (with first two generations of level-1 
quarks) associated 
productions as a function of $R^{-1}$ at the LHC energy 
of 14 TeV.\footnote{Later in this work, we demonstrate the prospect of a 10 TeV
LHC run which is imminent in a few years time after the scheduled take-off
in November, 2009.}
Also presented are the rates for the associated production of 
level-1 gluon along with level-1 bottom quarks and pair production of level-1
top quarks. We closely agree with the existing literature on these estimates
at appropriate limits.
From $g_1$-pair production 
and $g_1 b_1$ associated production processes, a combinatoric enhancement 
in the rate for inclusive single $H^\pm_1$ final state would be present since
both $g_1$ and $b_1$ can decay into $H^\pm_1$. Pair productions cross sections
of $b_1$ (including  $b_1^D \bar{b}_1^D$, $b_1^D \bar{b}_1^S + c.c.$, 
$b_1^D b_1^D + c.c.$) and $t_1^S$ are also indicated since they could directly
decay into level-1 charged Higgs bosons. While obtaining the
results presented in this plot we checked that the results of 
\cite{Macesanu:2002db,Bhattacherjee:2007wy} are reproduced under appropriate setups.
In Fig. \ref{fig:gluon1-decay} 
we present the decay 
branching fractions of  the level-1 KK gluon to the third generation level-1 
quarks, i.e., $b_1$ and $t_1$, as functions of $R^{-1}$. 
In Figs. \ref{fig:dt1dk} and \ref{fig:st1dk} similar variations of 
the subsequent decays of the level-1 lighter and heavier top quarks
($t_1^S$ and $t_1^D$), respectively to $H^\pm_1$ are illustrated.
A corresponding set of plots are  presented in Figs.\ref{fig:db1dk}
for the level-1 bottom quark, $b_1^D$.

\begin{figure}[hbtp]
\centering
\subfigure[]
{\label{fig:dt1dk}
\includegraphics[height=7.5cm, width= 5.7cm, angle =270]{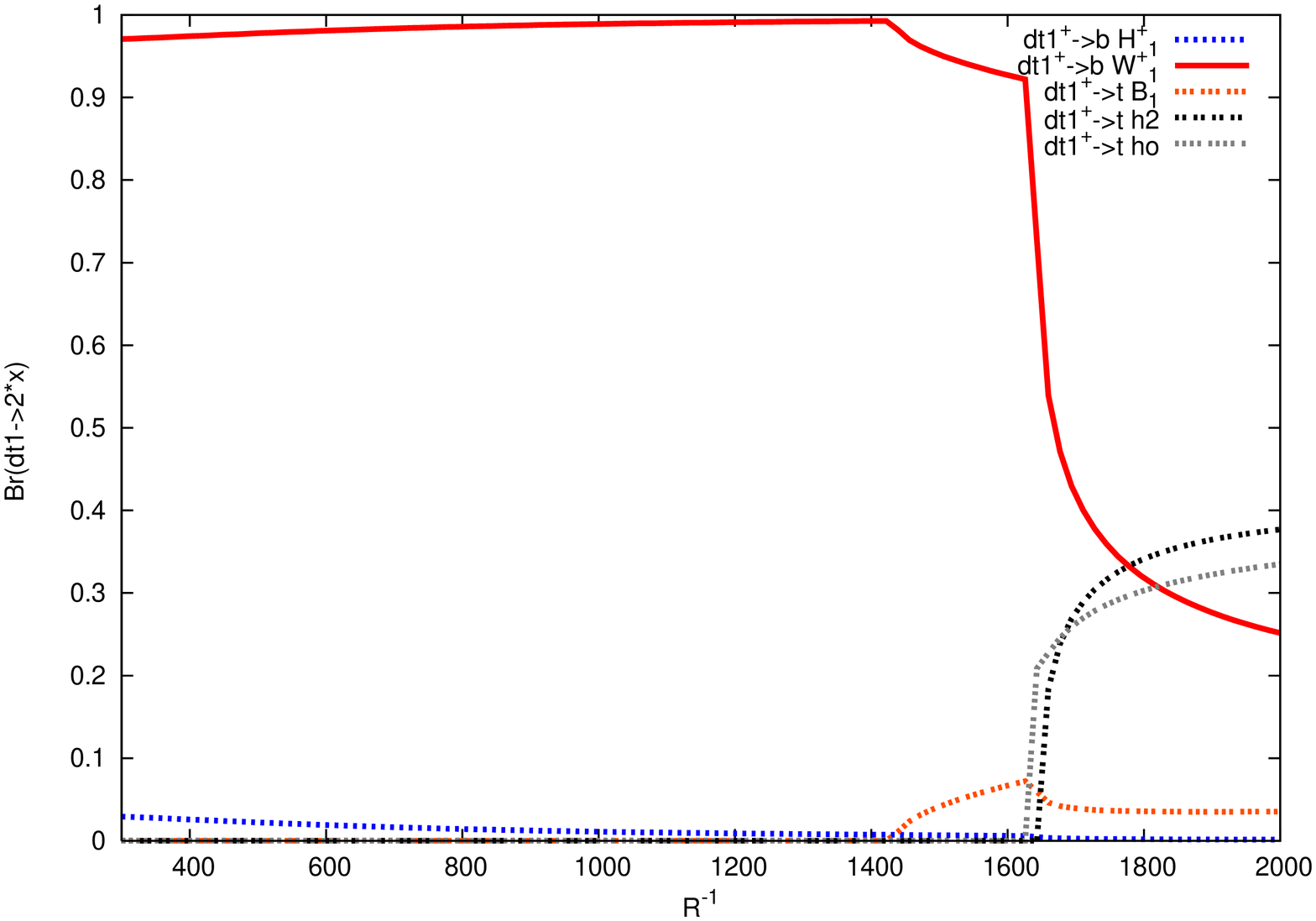}}
\hskip 15pt
\subfigure[]
{\label{fig:st1dk}
\includegraphics[height=7.5cm, width= 5.7cm, angle =270]{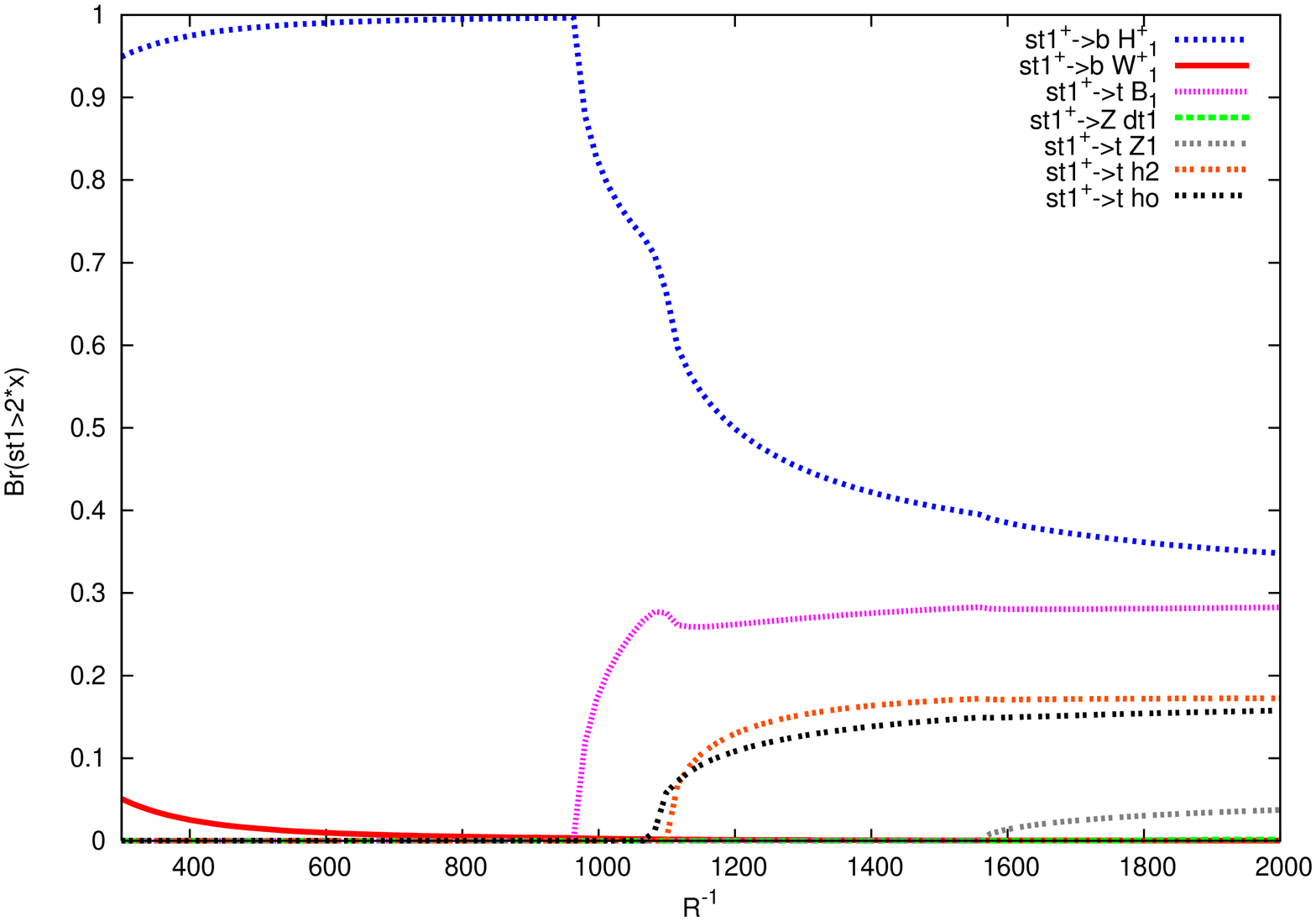}}
\caption{\small 
Variations of different decay branching fractions with respect to $R^{-1}$ (in GeV)
for (a) the lighter level-1 KK-top quark (predominantly doublet type) and
(b)  the heavier level-1 KK-top quark (predominantly singlet type).}
\end{figure}
\begin{figure}[hbtp]
\centering
\includegraphics[height=7.5cm, width= 5.7cm, angle =270]{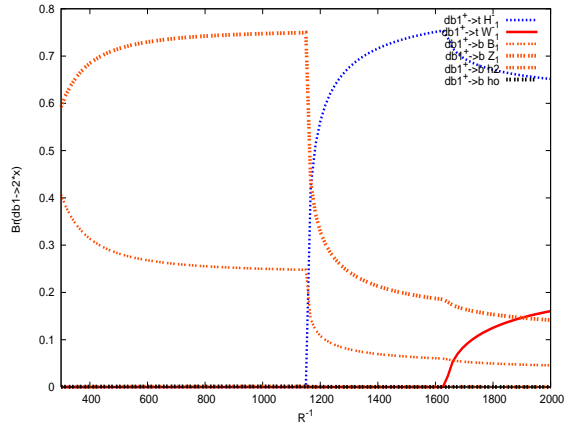}
\caption{\small
Variations of different decay branching fractions with respect to $R^{-1}$ (in GeV)
of the lighter (predominantly doublet type) level-1 KK-bottom quark.}
\label{fig:db1dk}
\end{figure}

From Fig. \ref{fig:strong-prod} it is apparent that a simple sum over the 
cross sections from different production processes contributing to a final 
state with a single $H^\pm_1$ could vary between 1000 pb and 10 fb for  
$300 < R^{-1} < 2000$ GeV.
This is a conservative estimate since many of the contributing processes
may have a combinatorial effect (as indicated above) which may lead to an
enhanced effective cross section.
 
From Fig. \ref{fig:gluon1-decay} it is seen that $g_1$ could always decay into
$b_1^D$ and $b_1^S$ (as far as third generation level-1 quarks are concerned, 
which may
subsequently decay to $H^\pm_1$). The decay $g_1 \to t_1^D t$ opens up only at a 
higher value of $R^{-1}$ around 1100 GeV when the mass-splitting between 
$g_1$ and $t_1^D$ 
allows for the top-mass. Once it opens up the corresponding branching fraction 
easily overtakes the ones for the decays to level-1 bottom quarks.
The reason behind this is the appreciable mixing in the level-1 
top-quark sector compared to that in the level-1 bottom quark sector.
As $m_{t_1^S} > m_{t_1^D}$, it is expected that
$g_1 \to t_1^S t$ opens up at an even larger value of $R^{-1}$. 
Note that, in any case, these branching fractions are only in the order of 
a few percent, the bulk fraction being shared by the decay modes to first 
two generations of quarks.

Figs. \ref{fig:dt1dk} and \ref{fig:st1dk} reveal that it is only $t_1^S$ 
that can have an appreciable branching fraction to $H^\pm_1$ reaching almost
100\% between $600 < R^{-1} < 950$ GeV. Even for $R^{-1}$ around 2 TeV the
branching fraction to $H_1^\pm$ remains appreciable ($\simeq 35\%$). It is
observed that the rest of the branching fraction is shared by 
$t_1^S$ decaying to level-1 neutral Higgs bosons or $B_1$ along with the 
SM top quark which are open for $R^{-1} \gtrsim 1$ TeV. 
In other words, the small mass of the associated SM $b$-quark increases the
possibility of an enhanced production of $H^\pm_1$ in the decay of $t_1^S$.
Note that, the chiral mixing being not so large even for the
level-1 quarks from the third generation, $t_1^S$ remains predominantly of 
the singlet type. Hence, its decay to $W^\pm_1$ is always suppressed. Just
the reverse is true for $t_1^D$, i.e., it dominantly decays to $W^\pm_1$ 
before being taken over by decays to level-1 neutral Higgs bosons at around
$R^{-1}=2$ TeV.
The decay patterns are consistent with the couplings involved whose forms
are rather complicated (see, for example, Appendix A of ref.\cite{Buras:2002ej}).

Thus, as far as production of $H^\pm_1$ is concerned, the
bottom line is that having $t_1^D$ would be of no use while $t_1^S$ could 
be a significant source (with a branching ratio $\gtrsim 35\%$) over the 
phenomenologically interesting range of $R^{-1}$.

In Fig. \ref{fig:db1dk} 
we present the decay branching
fractions of $b_1^D$.
Its decay to $H_1^\pm$ 
opens up for larger $R^{-1}$ ($\simeq 1150$ GeV) such that the mass-split 
between $b_1^D$ and $H_1^\pm$ allows for the mass of the accompanying top quark.
It grows fast and 
dominates for large $R^{-1}$, the corresponding branching fraction reaching 
up to 65\%. The effects are again consistent with the form of the involved couplings
mentioned above.
On the other hand, $b_1^S$ decays into $b B_1$ with almost 100\% branching 
fraction over the entire range of $R^{-1}$.

The summary of the information obtained from the above set of figures is the
following. Only $t_1^S$ and $d_1^B$ can have appreciable branching fractions 
to $H_1^\pm$ with the former having the branching fraction to $H_1^\pm$
ranging between 35\% and 100\% over the accessible range of $R^{-1}$. 
However, having $t_1^S$ in the decay of $g_1$ is somewhat less probable 
and only takes 
place only at larger values of $R^{-1}$. On the other hand, $b_1^D$ can be
obtained in the decay of $g_1$ (although with a branching fraction of only
about 10\%, on an average, for 300 GeV $< R^{-1} < 2000$ GeV) this could 
yield $H_1^\pm$ under cascades for $R^{-1} > 1150$ GeV. Thus, it appears, 
that under cascade decays of gluino, $b_1$ induced $H^\pm_1$ production 
would exceed the one induced by $t_1$. 

\begin{figure}[hbtp]
\centering
\subfigure[]
{\label{fig:cascade10}
\includegraphics[height=7.5cm, width= 5.7cm, angle =270]{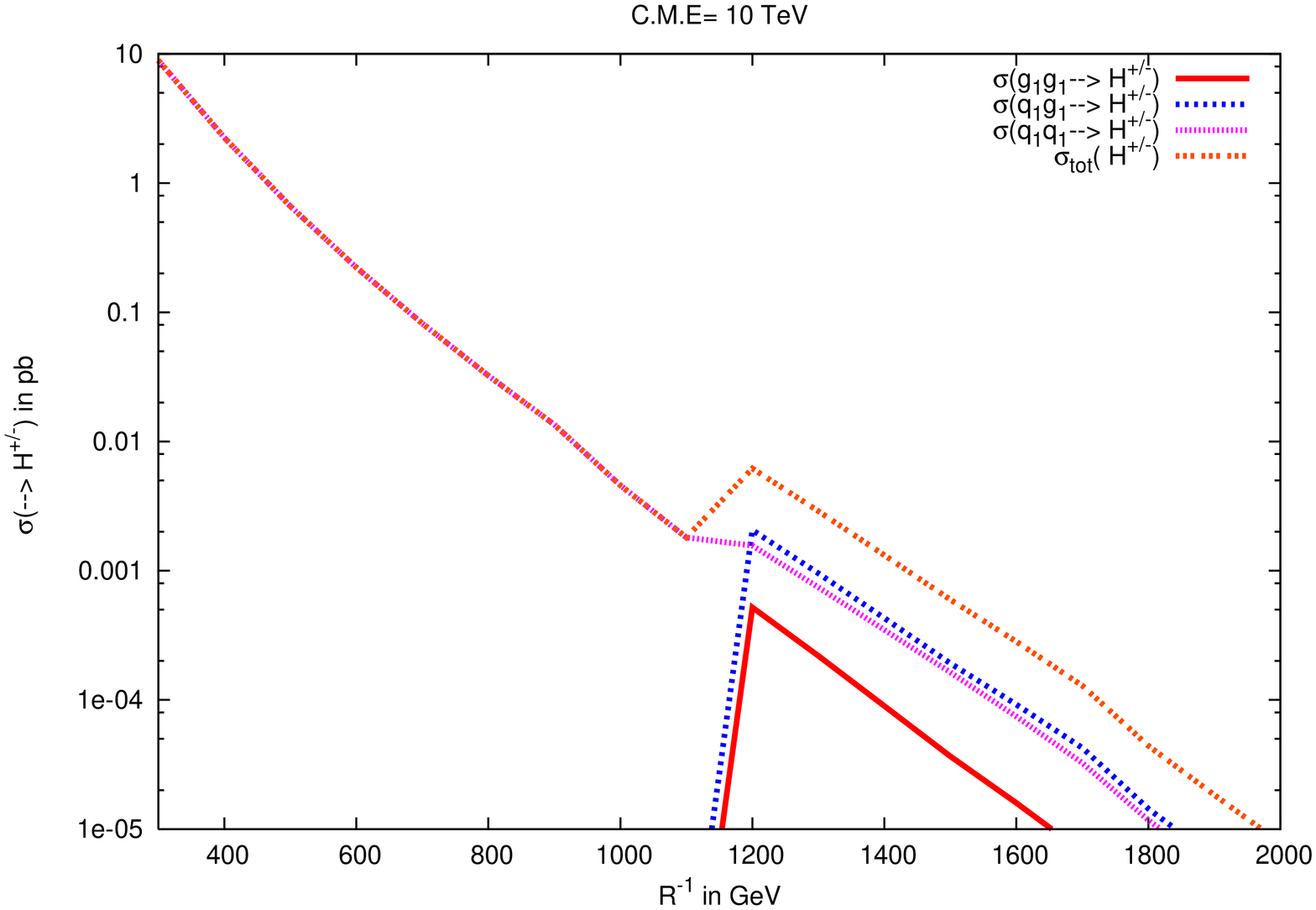}}
\hskip 15pt
\subfigure[]
{\label{fig:cascade14}
\includegraphics[height=7.5cm, width= 5.7cm, angle =270]{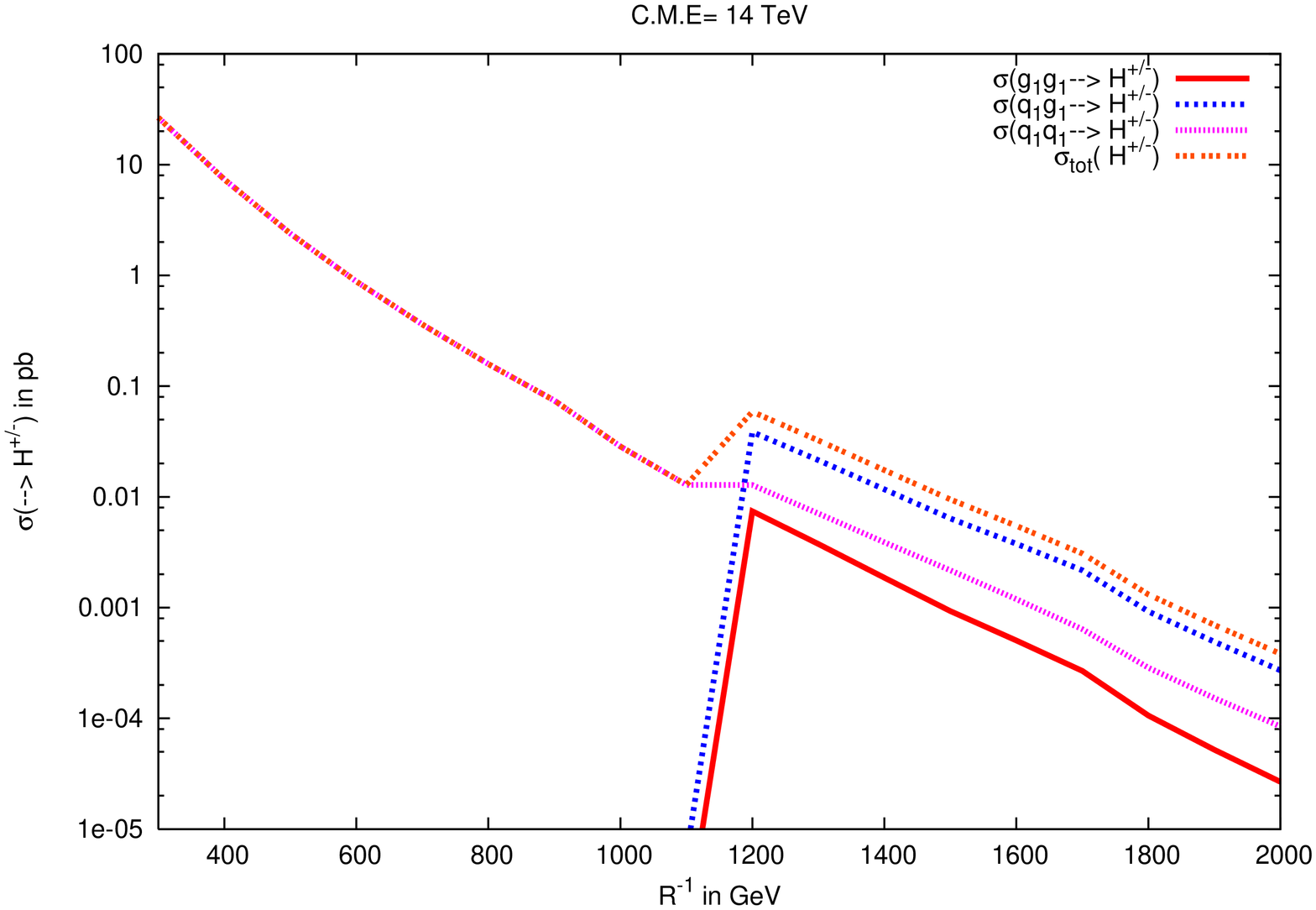}}
\caption{\small
Variations of effective production cross sections of an inclusive 
KK level-1 charged 
Higgs bosons under cascade decays of UED excitations
as functions  of $R^{-1}$.
}
\end{figure}

In Figs. \ref{fig:cascade10} and \ref{fig:cascade14} we compare the total 
contributions to $H^\pm_1$
production from these cascades and that from the direct productions of $t_1$ 
and $b_1$ at the LHC for centre of mass energies of 10 and 14 TeV respectively. 
Up to about 1.1 TeV, the sole contribution to $\h1pm$ comes dominantly 
from the production
of $t_1^S \bar{t}_1^S$-pair and their decays. As mentioned earlier, pair production of $b_1$ quarks
could not play a role here since $\h1pm$-s from their decays would always be
accompanied by top quarks and the mass-split between $b_1$ and $\h1pm$ is not
simply enough to accommodate the same for such low values of $\rinv$. 
As pointed out earlier, contribution to $\h1pm$ final states from the cascades 
of level-1 KK gluon turns on as soon as $\rinv$ is large enough 
($\gtrsim 1.1$ TeV) such that $g_1 \to t_1^S \bar{t}$ opens up.
There, as is clear from these figures, the contributions from the cascade decays 
could be appreciable and may enhance the yield of $\h1pm$ by an 
order of magnitude for both 10 and 14 TeV runs. 
The observation is significant in view of 
not so big a rate for $H^\pm_1$ via direct productions 
of the level-1 $b$ and $t$ quarks. It has already been pointed out in
earlier sections that detecting $H^\pm_1$ at the LHC could be a difficult
task. This is because the decay products of $H^\pm_1$ can be rather soft 
and thus, an experiment like the LHC could turn out to be rather insensitive 
to them. Unless we may expect
a radical improvement in the sensitivity of the experiment to softer 
decay products, the only way we could overcome this limitation is by having 
a larger rate for $H_1^\pm$ in the first place. This work advocates that
the production of $H^\pm_1$ through UED-cascades could come into aid and may
somewhat compensate for the poor sensitivity over a favourable
range of $R^{-1}$. Interestingly enough, such an enhancement comes to our aid
for heavier masses (higher values of $\rinv$) only thus could extend the reach
of the LHC by 250--300 GeV (by looking at the effective cross sections) for both
10 and 14 TeV runs. On the other hand,
it can be seen from these figures that a 14 TeV run would match the yield from
a 10 TeV run with masses heavier by 300-400 GeV. However, the absolute yields
are predicted to be quite small for the 10 TeV run. For example, requiring a
minimum of 100 raw $\h1pm$ events before folding in the usual suppression factors
like the lepton (particularly, $\tau$) reconstruction efficiency, trigger-related factors and 
other kinematic cuts, the tentative reach in $\rinv$ at two different
center-of-mass energies as a function of accumulated luminosities are as follows:
\vskip 5pt
\begin{table}[h]
\begin{center}
\begin{tabular}{||c||c|c|c|c||}
\hline
\hline
$\int {\cal L}\, dt$ (fb$^{-1}$) & 1 & 10 & 50 & 100 \\
\hline
\hline
$\sqrt{s}=10$ TeV & 600 & 1000  & 1200  & 1400 \\
\hline
$\sqrt{s}=14$ TeV & 900 & 1400 & 1500  &  1750 \\
\hline
\hline
\end{tabular}
\end{center}
\caption{\small Tentative reaches in $\rinv$ in GeV (by requiring 100 events
at the level of effective production of $\h1pm$ from Figs. \ref{fig:cascade10} 
and \ref{fig:cascade14})
as a function of accumulated luminosity for 10 TeV and 14 TeV 
runs of the LHC.}
\end{table}
\noindent
This means that the 14 TeV machine clearly has an edge over the 10 TeV option
as far as reach in $\rinv$ is concerned. However, as we can see from the Table 1,
a moderate volume of data (10 fb$^{-1}$) could be what is needed at a 10 TeV LHC 
run to probe $\rinv \sim 1$ TeV. This is already encouraging.

It is also to be noted that reaching out to $\rinv\approx 1.4$ TeV cannot be
a smooth exercise. For, as pointed out in section 3, with the SM Higgs mass
set at 120 GeV, approaching $\rinv \approx 1.4$ TeV from the lower side would
inevitably make $\h1pm$ the NLKP. This can make $\h1pm$ long-lived at the 
colliders and its signatures could be tantalizing enough.

\section{Summary and outlook}
We demonstrate that cascade decays of level-1 KK excitations of the MUED 
scenario could be a significant source of level-1 charged Higgs boson, 
$\h1pm$, of
the scenario. It is pointed out that cascades contribute to the yield of
$\h1pm$ significantly only beyond a certain threshold value of $\rinv$
which is instrumental in generating the required mass-splitting between the
level-1 KK gluon ($g_1$) and the level-1 top quark ($t_1$) such that $g_1$
produced in hard scattering could decay into $t_1$ which in turn cascades 
down to $\h1pm$. It is shown that with this added contribution to the yield
of $\h1pm$, the reach in $\rinv$ could be considerably enhanced, both at
10 TeV and 14 TeV runs with moderate amount of accumulated luminosity.
However, some of the generic experimental challenges with possible signals of
$\h1pm$ were also highlighted. 
A detailed generator-level analysis of the MUED Higgs sector
including major detector effects is beyond the scope of the present
work and can be discussed elsewhere.

The importance of detecting  the level-1 MUED Higgs bosons should be 
put in perspective. It is clear that the spectra of level-1 UED Higgs 
bosons is similar to that of the Higgs bosons of the MSSM with the 
heavier Higgs bosons of the MUED being rather heavy 
($\sim {\cal O}$(TeV)) and degenerate (among themselves and also with 
other low-lying excitations of the MUED from the same KK-level). 
With the LHC running, a situation can be envisaged where one finds only
one Higgs boson and no other resonances of a plausible new physics 
scenario (the so-called `lone Higgs' scenario \cite{Hsieh:2008jg}) and one may 
like to identify this Higgs boson as that  of the SM or the one from 
one of the scenarios beyond the same, like the MSSM, MUED or the Little 
Higgs scenario with conserved $T$-parity (LHT). It was demonstrated in 
ref. \cite{Hsieh:2008jg} that, for example, the size and the sign of deviation 
in the rate for $gg \to h \to \gamma \gamma$ from the SM expectation 
could be an indicator of the nature of the new physics. However, as long
as the deviation at a given point of time remains small (say, below 30\%)
it would not qualify to be a robust determinant, at least during the early
stages of the experiment.

On the other hand, given the reach of the LHC can reasonably be expected to be in 
the same ballpark for both MSSM and the MUED, one could envisage an alternate 
possibility where one sees not only the Higgs boson but also some
other MSSM-like excitations ($\sim \cal{O}$(TeV)). However, it may turn out that
the corresponding deviation from the SM expectation in the 
$gg \to h \to \gamma \gamma$ mode is not convincing enough for whatsoever reason.
Such a situation could appear much confounding for two reasons:
(i) the well-known SUSY-UED confusion for which an immediate identification
of the observed resonances in favour of one or the other of these 
contending scenarios
may not be possible,
(ii) when the heavier Higgs bosons of the MSSM could still 
have showed up, in principle
(since they could still be drastically lighter), even if the observed 
resonances (like the partners of strongly interacting quarks and gluon) are 
somewhat heavy (though observable), we are observing only one Higgs boson. This is 
not the case for MUED where the mass-scale of all the UED-excitations 
including the heavier Higgs bosons is determined by $R^{-1}$.
An observation of this kind, i.e., non-observation of heavier Higgs bosons
up to masses close to $R^{-1}$ that corresponds to the masses of the other 
observed resonances, may favour a scenario like the MUED, of
course, without in any way, excluding the possibility of a SUSY like scenario,
as such. 

Under such circumstances, one may require to find, simultaneously, the 
corresponding neutral Higgs bosons of the emerging scenario in addition
to the charged Higgs boson. For a given
$R^{-1}$, their respective yields are expected to be somewhat smaller than
the corresponding charged Higgs boson. On top of this, the neutral Higgs 
excitations being close in mass and being expected to be observed in a
similar way, their individual resolutions are likely 
to get affected in any experimental setup. It is thus clear that nailing 
them down at the LHC would not be an easy task though could turn out to be
a crucial one.

It may be reiterated that not only the Higgs sector of the MUED but also
the whole spectrum of the scenario  is rather degenerate at a given KK-level. 
Surely,
the degeneracy is much severe and of a serious nature in the Higgs sector 
from the point of view of their detectability at a future collider. While
the phenomenological implications of the MUED is worth studying in its own
right (perhaps with a status similar to that enjoyed by the mSUGRA framework in SUSY 
studies, i.e., as a benchmark 
scenario with a somewhat tractable number of free parameters) its viability
as a realistic framework should not be over-stressed. However, in the spirit
of revealing and exploring diverse phenomenological possibilities at the
future colliders in terms of novel and unexpected signatures and in the form of
viable scenarios, such studies are very much called for and are being much
advocated \cite{Strassler:2008jq}.
On the other hand, in recent times, 
studies revealed that even the framework of MUED could accommodate several 
alternate candidates for the LKP, like the hypercharge gauge boson $B_1$, 
the charged Higgs boson $H_1^\pm$ and even the level-1 KK graviton $G_1$ 
\cite{Cembranos:2006gt} and thus potentially has the seeds that can lead to
diverse phenomenological situations.
There have also been attempts to obtain some robust, non-minimal realisations 
of the MUED \cite{Flacke:2008ne} by including non-zero 
boundary-localized terms at the cutoff scale $\Lambda$ for a part of the UED 
spectrum. Frameworks like these could enrich the resulting UED-phenomenology
at future colliders, including that of the Higgs sectors of such scenarios. 

\vskip 30pt
\noindent
{\Large \bf Acknowledgements}

\vskip 10pt
\noindent
The authors like to thank Bogdan Dobrescu, Monoranjan Guchait, 
Graham Kribs, Anirban Kundu,
Andreas Nyffeler and Ashoke Sen for insightful discussions at different
stages of the work. The work of BB has been supported by a research
fellowship from the University Grants Commission, Government of India.
BB thanks the Regional Centre for Accelerator-based Particle Physics 
(RECAPP), HRI (funded by the Department of Atomic Energy, Government of 
India under the XIth 5-year Plan) for a Young Associateship. 
AD thanks the Theory Division of CERN, Switzerland for their hospitality
during the summer of 2009 when part of the present work was carried out. 
Computation 
required for this work are done on the computing facilities of RECAPP and 
the Cluster Computing Facility at HRI {\tt (http://cluster.hri.res.in)}.


%
\end{document}